\documentclass[twocolumn]{aastex631}

\usepackage{amsmath}
\usepackage{graphicx}
\usepackage{longtable}
\usepackage{multirow}
\usepackage{subfigure}
\usepackage{epstopdf}

\shorttitle{NS-NS and NS-BH from BdHNe}
\shortauthors{Becerra et al.}

\begin{document}
\title{On the formation of compact-object binaries from binary-driven hypernovae}

\author[0000-0002-3262-5545]{L.~M.~Becerra}
\affiliation{Escuela de F\'isica, Universidad Industrial de Santander, A.A.678, Bucaramanga, 680002, Colombia }
\affiliation{ICRANet, Piazza della Repubblica 10, I-65122 Pescara, Italy}

\author[0000-0003-2624-0056]{C.~L.~Fryer}
\affiliation{Center for Theoretical Astrophysics, Los Alamos National Laboratory, Los Alamos, NM, 87545, USA}
\affiliation{Computer, Computational, and Statistical Sciences Division, Los Alamos National Laboratory, Los Alamos, NM, 87545, USA}
\affiliation{The University of Arizona, Tucson, AZ 85721, USA}
\affiliation{Department of Physics and Astronomy, The University of New Mexico, Albuquerque, NM 87131, USA}
\affiliation{The George Washington University, Washington, DC 20052, USA}

\author[0000-0003-4904-0014]{J. A. Rueda}
\affiliation{ICRANet, Piazza della Repubblica 10, I-65122 Pescara, Italy}
\affiliation{ICRANet-Ferrara, Dip. di Fisica e Scienze della Terra, Universit\`a degli Studi di Ferrara, Via Saragat 1, I--44122 Ferrara, Italy}
\affiliation{ICRA, Dipartamento di Fisica, Sapienza Universit\`a  di Roma, Piazzale Aldo Moro 5, I-00185 Rome, Italy}
\affiliation{Department of Physics and Earth Science, 
University of Ferrara, Via Saragat 1, I-44122 Ferrara, Italy}
\affiliation{INAF, Istituto di Astrofisica e Planetologia Spaziali, Via Fosso del Cavaliere 100, 00133 Rome, Italy}

\author[0000-0003-0829-8318]{R.~Ruffini}
\affiliation{ICRANet, Piazza della Repubblica 10, I-65122 Pescara, Italy}
\affiliation{ICRA, Dipartamento di Fisica, Sapienza Universit\`a  di Roma, Piazzale Aldo Moro 5, I-00185 Rome, Italy}
\affiliation{Universit\'e de Nice Sophia-Antipolis, Grand Ch\^ateau Parc Valrose, Nice, CEDEX 2, France}
\affiliation{INAF, Viale del Parco Mellini 84, 00136 Rome, Italy}

\begin{abstract}
We present smoothed-particle-hydrodynamics (SPH) simulations of the binary-driven hypernova (BdHN) scenario of long gamma-ray bursts (GRBs), focusing on the binary stability during the supernova (SN) explosion. The BdHN progenitor is a binary comprised of a carbon-oxygen (CO) star and a neutron star (NS) companion. The core collapse of the CO leads to an SN explosion and a newborn NS ($\nu$NS) at its center. Ejected material accretes onto the NS and the $\nu$NS. BdHNe of type I have compact orbits of a few minutes, the NS reaches the critical mass, forming a black hole (BH), and the energy release is $\gtrsim 10^{52}$ erg. BdHNe II have longer periods of tens of minutes to hours; the NS becomes more massive, remains stable, and the system releases $\sim 10^{50}$--$10^{52}$ erg. BdHN III have longer periods, even days, where the accretion is negligible, and the energy released is $\lesssim 10^{50}$ erg. We assess whether the system remains gravitationally bound after the SN explosion, leading to an NS-BH in BdHN I, an NS-NS in BdHN II and III, or if the SN explosion disrupts the system. The existence of bound systems predicts an evolutionary connection between the long and short GRB populations. We determine the binary parameters for which the binary remains bound after the BdHN event. For these binaries, we derive fitting formulas of the numerical results for the main parameters, e.g., the mass loss, the SN explosion energy, orbital period, eccentricity, center-of-mass velocity, and the relation between the initial and final binary parameters, which are useful for outlined astrophysical applications.
\end{abstract}

\section{Introduction}

The binary-driven hypernova (BdHN) model is born to satisfy the increasing demand from the stellar evolution and GRB theory and GRB observations for a binary progenitor of long GRBs \citep{1999ApJ...526..152F, 2012ApJ...758L...7R}. In particular, the fact that long GRBs are temporally and spatially associated with supernovae (SNe), moreover of type Ic \citep{2006ARA&A..44..507W, 2011IJMPD..20.1745D, 2012grb..book..169H}, that the energetics of GRBs and SNe are widely different, and that most massive stars belong to binaries \citep{2007ApJ...670..747K, 2012Sci...337..444S}. We refer to  \citet{2015PhRvL.115w1102F, 2019Univ....5..110R, 2022PhRvD.106h3002B, 2023ApJ...955...93A} for discussions on this topic.

The GRB progenitor is a CO-NS binary at the end of the thermonuclear evolution of the CO star, i.e., at the second core-collapse SN event in the evolution of the binary (the first SN formed the NS companion). Thus, the CO star explodes as an SN in the presence of the NS companion \citep{2012ApJ...758L...7R}. The different emissions observed in the GRB are explained by the sequence of physical processes triggered by the SN explosion \citep[see, e.g.,][]{2012ApJ...758L...7R, 2012A&A...548L...5I, 2014ApJ...793L..36F, 2015PhRvL.115w1102F, 2015ApJ...812..100B, 2016ApJ...833..107B, 2019ApJ...871...14B, 2022PhRvD.106h3004R, 2022PhRvD.106h3002B, 2023ApJ...955...93A}. The gravitational collapse of the iron core of the CO star forms a newborn NS ($\nu$NS) at its center, ejecting the pre-SN stellar outer layers. The ejecta triggers an accretion process onto the NS companion and the $\nu$NS. Fallback accretion causes the latter. Such accretion processes proceed at hypercritical (i.e., highly super-Eddington) rates as the gravitational energy gain is mostly taken away by the emission of MeV-neutrinos \citep{2016ApJ...833..107B, 2018ApJ...852..120B}. In compact binaries with a few minutes orbital periods, the hypercritical accretion onto the NS companion brings the critical mass, which induces its gravitational collapse and forms a rotating (Kerr) BH. These systems are called BdHN I and are the most energetic with an energy release $\gtrsim 10^{52}$ erg, like GRBs 130427A \citep{2019ApJ...886...82R}, 180720B \citep{2022ApJ...939...62R}, and 190114C \citep{2021PhRvD.104f3043M, 2021A&A...649A..75M}. We refer to \citet{2021MNRAS.504.5301R} for details on the general analysis of 380 BdHNe I. In less compact binaries with periods from tens of minutes to hours, the NS companion does not reach the critical mass and holds stable as a more massive and fast-rotating NS. These systems, called BdHN II, release energies in the range $\sim 10^{50}$--$ 10^{52}$ erg. An example is GRB 190829A \citep{2022ApJ...936..190W}. Wide CO-NS binaries with periods of up to days are called BdHN III and release $\lesssim 10^{50}$ erg. An example is GRB 171205A \citep{2023ApJ...945...95W}.

Possible binary evolution paths for forming the CO-NS binaries of BdHNe have been discussed in \citet{2015PhRvL.115w1102F}. The most likely evolutionary path appears to be that of ultra-stripped binaries, which have been introduced to explain the population of the Galactic NS-NS binaries and low-luminosity and/or rapid-decay-rate SNe \citep{2013ApJ...778L..23T, 2015MNRAS.451.2123T}. The evolution starts from two massive stars. The first core-collapse SN of the primary star forms an NS. After that, the system undergoes a series of mass transfer phases, ejecting the secondary's hydrogen and helium shells to produce a binary composed of a massive CO star and an NS. Therefore, in this picture, it has been assumed that at the second SN explosion in the binary evolution, when the iron core of the He/WR/CO star undergoes gravitational collapse, an NS-NS forms \citep{2015MNRAS.451.2123T, 2017ApJ...846..170T, 2018Sci...362..201D}. As we anticipated, the BdHN model explores the possibility that the NS companion of the CO star, for short orbital periods, can accrete enough mass from the SN ejecta to reach the critical mass and form a Kerr BH. Numerical simulations that include the most relevant physical processes occurring in the cataclysmic event have shown the occurrence of this phenomenon \citep{2016ApJ...833..107B, 2019ApJ...871...14B, 2022PhRvD.106h3002B}. Therefore, the numerical simulations show three possible fates of a BdHN: forming an NS-NS (in BdHN II), an NS-BH (in BdHN I), or two runaway NSs (mostly in BdHN III).

If the binaries remain gravitationally bound, they are expected to merge through the emission of gravitational waves, leading to short GRBs. Hence, BdHNe events establish a direct connection between short and long GRBs. Such a connection is also supported by their occurrence rates \citep[see, e.g.,][for details]{2016ApJ...832..136R, 2018ApJ...859...30R, 2023arXiv230605855B}. For instance, the observed (isotropic) density rate of BdHNe I in the local universe is low, $\sim 1$ Gpc$^{3}$ yr$^{-1}$ \citep[see, e.g.,][]{2016ApJ...832..136R, 2018ApJ...859...30R}, which is consistent with the short orbital periods required in BdHNe. Long GRBs are rare astrophysical sources. Thus, these CO-NS binaries could be a small subset of the ultra-stripped binaries. \citet{2015PhRvL.115w1102F} estimated that only $\sim 0.1\%$-–$1\%$ of ultra-stripped binaries is required to explain the BdHN I population. 

Therefore, estimating the conditions under which the SN explosion disrupts the systems and which form NS-BH or NS-NS out of a BdHN event is crucial to set and predict the evolutionary relation between the long and short GRBs \citep{2016ApJ...832..136R, 2018ApJ...859...30R}. In this paper, we perform three-dimensional (3D), smoothed-particle-hydrodynamics (SPH) simulations of the SN explosion and the subsequent expansion of the ejected material under the gravitational field of the NS companion and the $\nu$NS, focusing on determining whether the system holds bound or not, and establish the binary and explosion features leading to the various fates.

This article is structured as follows. In Section \ref{sec:sims}, we discuss the SPH simulations of BdHNe and provide details about the initial configurations. In Section \ref{sec:results}, we present the results of exploring the initial parameter space, including the initial binary period (Section \ref{sec:Pinit}), the initial SN energy and mass of the NS companion (Section \ref{sec:SNenergy}), and the progenitor of the CO core (Section \ref{sec:COprog}). Finally, we discuss our results in section~\ref{sec:clonclusions}.

\section{Simulations}\label{sec:sims}
%
\begin{table}
  \centering
\begin{tabular}{c|ccccc}
      N & $M_{\rm \nu ns,f}$ & $M_{\rm ns,f}$ & $a_{\rm orb,f}$ & $P_{\rm orb,f}$ & $e_f$\\
    Millions & $M_\odot$ & $M_\odot$  & $10^{10}$~cm & min &     \\ \hline
          1      & $1.970$ & $2.086$ & $7.862$  & $99.48$  & $0.856$   \\          

          2      & $1.949$ & $2.072$ & $9.357$  & $129.72$ & $0.880$     \\                 
        
          3      & $1.941$ & $2.061$ & $10.593$ & $156.64$ & $0.889$    \\                 
         
          4      & $1.933$ & $2.058$ & $10.376$ & $152.06$ & $0.899$   \\   

          5      & $1.930$ & $2.056$ & $10.453$ & $151.13$ & $0.895$      \\                 \hline
               \end{tabular}
               \caption{ Convergence test: Final binary parameters for an initial binary system formed by a $2\,M_\odot$ NS and a CO evolving from a progenitor with $M_{\rm zams}=25M_\odot$ in an initial orbital period of $\sim5$~min.}
        \label{tab:conv}
\end{table}

We used the \textsc{SN-SPH} code  \citep{2006ApJ...643..292F} to follow the evolution of the binary stars (the $\nu$NS and the NS companion) as the material ejected from the SN explosion expands. The initial setup has been described in \cite{2019ApJ...871...14B}. Here, we briefly summarize it.

The SPH simulation starts when the SN shock front reaches the outer radius of the CO, i.e., we built the initial conditions for the SN ejecta, mapping to a 3D-SPH configuration, the 1D core-collapse SN simulation of \cite{2018ApJ...856...63F}. We then add two point masses to the simulations to model the gravitational effects of the $\nu$NS and the NS companion. These particles only interact gravitationally with the SN ejecta particles and between each other. Moreover, according to the algorithm described in \cite{2019ApJ...871...14B}, they can accrete other particles from the SN ejecta.

The code tracks the evolution of the position and velocity of all the SN ejecta particles and the point-mass particles. We calculate the total energy of the $\nu$NS and the NS companion as the sum of the total kinetic energy, $E_{\rm kin}$, relative to the binary center of mass, and the  gravitational energy between the two stars, $E_{\rm grav}$, i.e.,
\begin{equation}\label{eq:Energy_orb}
    E_{\rm tot}= \frac{1}{2}\frac{M_{\rm \nu ns}M_{\rm ns}}{M_{\rm \nu ns}+M_{\rm ns}} |\vec{v}_{\rm \nu ns}-\vec{v}_{\rm ns}|^2 -\frac{GM_{\rm \nu ns}M_{\rm ns}}{|\vec{r}_{\rm \nu ns}-\vec{r}_{\rm ns}|}\, ,
\end{equation}
where ($M_{\rm \nu ns},\vec{r}_{\rm \nu ns},\vec{v}_{\rm  \nu ns})$ and $(M_{\rm ns}, \vec{r}_{\rm  ns},\vec{v}_{\rm  ns})$ are the mass, vector position and velocity of the $\nu$NS and the NS companion, respectively.
Suppose the $\nu$NS and the NS companion are bound, and the system energy is negative, i.e., $E_{\rm tot}<0$. Therefore, the orbital separation can be determined by
\begin{equation}\label{eq:aorb}
    E_{\rm tot}= -\frac{1}{2}\frac{GM_{\rm \nu ns}M_{\rm ns}}{a_{\rm orb}}.
\end{equation}
The orbital period, $P_{\rm orb}$, follows from the Kepler's law
\begin{equation}\label{eq:Porb}
    \left(\frac{2\pi}{P_{\rm orb}}\right)^2 = \frac{G(M_{\rm \nu ns}+M_{\rm ns})}{a_{\rm orb}^3}\, ,
\end{equation}
while the orbit eccentricity, $e$, is
\begin{equation}\label{eq:ecc}
    |\vec{L}_{\rm tot}| = M_{\rm \nu ns}M_{\rm ns}\sqrt{\frac{Ga_{\rm orb}(1-e^2)}{M_{\rm \nu ns}+M_{\rm ns}} }\, , 
\end{equation}
with $\vec{L}_{\rm tot}$ the total angular momentum of the binary system stars relative to its center of mass. If the $\nu$NS and the NS companion become unbound, the system's total energy is positive, i.e., $E_{\rm tot}>0$.

To determine the total number of particles needed in the simulations, we tested the convergence of the final orbital parameters (orbital period,  binary separation, and eccentricity) with the total number of SPH particles. We performed this analysis for an initial binary system consisting of a $2~M_\odot$ NS and a CO evolved from a ZAMS progenitor of $M_{\rm zams}=25~M_\odot$, with an initial orbital period of $5$~min. The SPH simulations stop when the ejecta mass gravitationally bound to the stars ($\nu$NS and NS companion) is less than $10^{-3}~M_\odot$, i.e., when the gravitational effect of the SN ejecta mass on the system evolution becomes negligible. 

For the present case, the final system remains gravitationally bound ($E_{\rm tot, final}< 0 $). Table~\ref{tab:conv} summarizes the results of these simulations.  We used a total SPH particle number, $N$, from 1 million to 5 million. We registered  the final masses of the $\nu$NS, $M_{\rm \nu ns,f}$, and the NS companion, $M_{\rm ns,f}$, the final orbital separation, $a_{\rm orb,f}$, the final binary period, $P_{\rm orb,f}$ and the final eccentricity, $e_f$.  As the number of particles in the simulations increases, the mass accreted by the $\nu$NS and the NS companion slightly decreases during the final orbital separation, orbital period, and eccentricity increase. Based on these results, we set a total particle number of $4$ million for all the simulations in this article. Throughout the article, we explore the results for three different ZAMS progenitors of the CO star, which we summarize in Table \ref{tab:ProgSN}.

\begin{table}
\centering
\setlength{\tabcolsep}{5pt}
\begin{tabular}{c|cccc}
  $M_\mathrm{ZAMS}$ & $M_\mathrm{rem} $ & $M_\mathrm{ej}$ &      $R_{\rm core}$        &      $R_{\rm star}$  \\
   $(\,M_\odot\,)$  & $(\,M_{\odot}\,)$ & $(\,M_\odot\,)$ & $(\,10^8\, \mathrm{cm}\,)$ & $(\,10^9\,\mathrm{cm}\,)$\\
   \hline
    $25$ & $1.85$ & $4.995$ & $2.141$ & $5.855$\\
    $30$ & $1.75$ & $7.140$ & $28.33$ & $7.830$\\
    $40$ & $1.85$ & $11.50$ & $19.47$  &$6.529$\\
\hline
\end{tabular}
\caption{Properties of the CO progenitors. The mass of the evolved CO star is given by the mass of the central remnant, $M_{\rm rem}$, and that of the outer layers, which will be the ejected mass, $M_{\rm ej}$, so $M_{\rm CO} = M_{\rm rem}+M_{\rm ej}$. The mass of the $\nu$NS is $M_{\nu \rm NS} = M_{\rm rem}$. The radius of the CO iron core is $R_{\rm core}$, and the total radius of the pre-SN star is $R_{\rm star}$. The progenitors are evolved with the KEPLER stellar evolution code \citep{2010ApJ...724..341H} and exploded using the 1D core-collapse code presented in \citet{1999ApJ...516..892F}.}
\label{tab:ProgSN}
\end{table}

\section{Results}\label{sec:results}
 
 %
\begin{figure}
    \centering
    \includegraphics[width=0.99\columnwidth,clip]{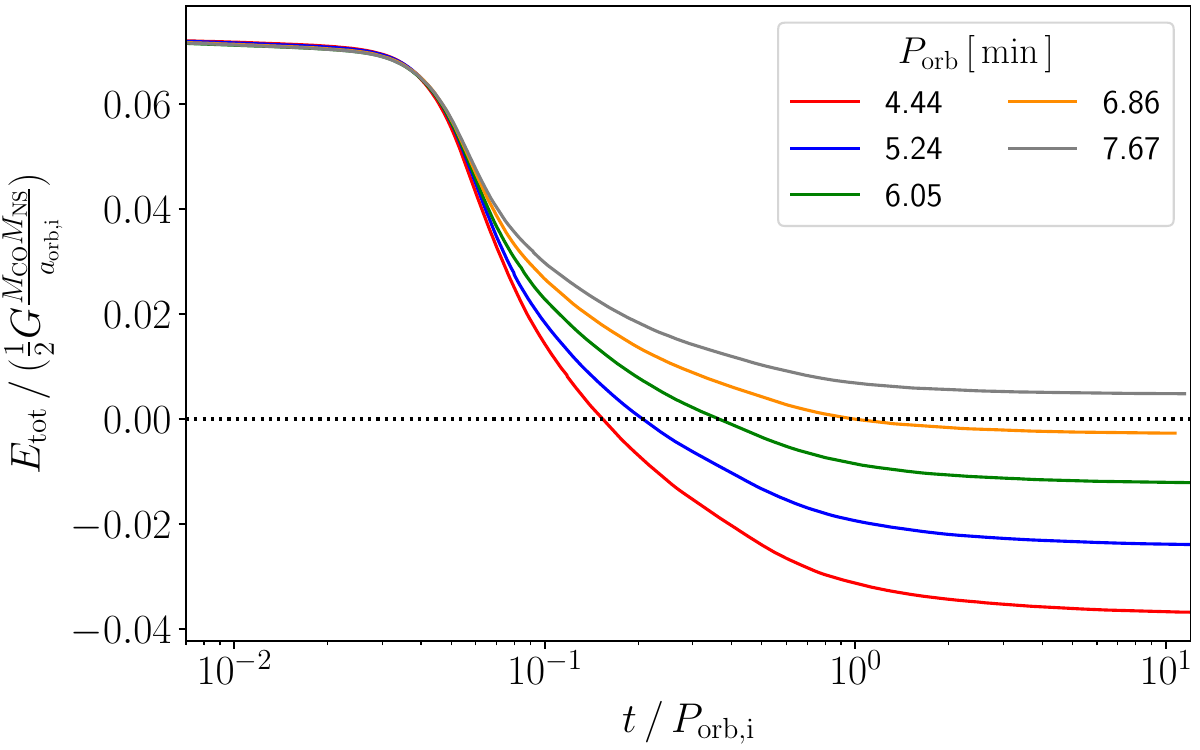} 
    \caption{Time evolution of the total energy of the $\nu$NS-NS system after the SN explosion for different initial orbital periods. The initial binary system was formed by an NS companion with $M_{\rm ns}=2~M_\odot$ and a CO with $M_{\rm CO}=6.8~M_\odot$, which leaves a $\nu$NS with $M_{\rm \nu NS}=1.85~M_\odot$ after its collapse. The black horizontal dotted line separates the systems that remain gravitationally bound ($E_{\rm tot}<0$) from the ones that do not ($E_{\rm tot}>0$). The total energy has been normalized to the absolute values of the pre-SN, CO-NS binary energy. 
    }
    \label{fig:Etot_Evol}
\end{figure}
\begin{figure*}
    \centering
    \includegraphics[width=0.4\textwidth,clip]{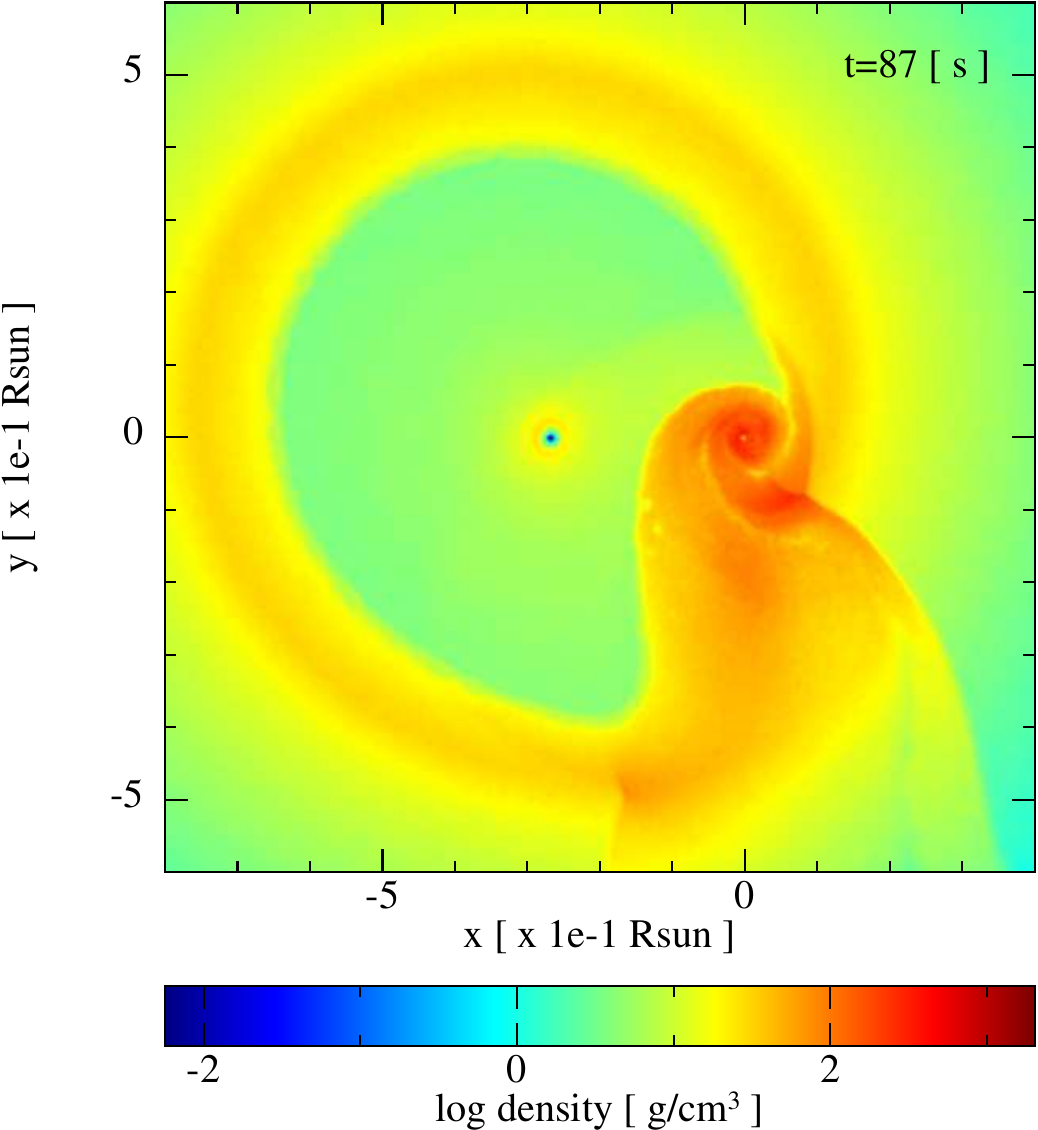} \includegraphics[width=0.4\textwidth,clip]{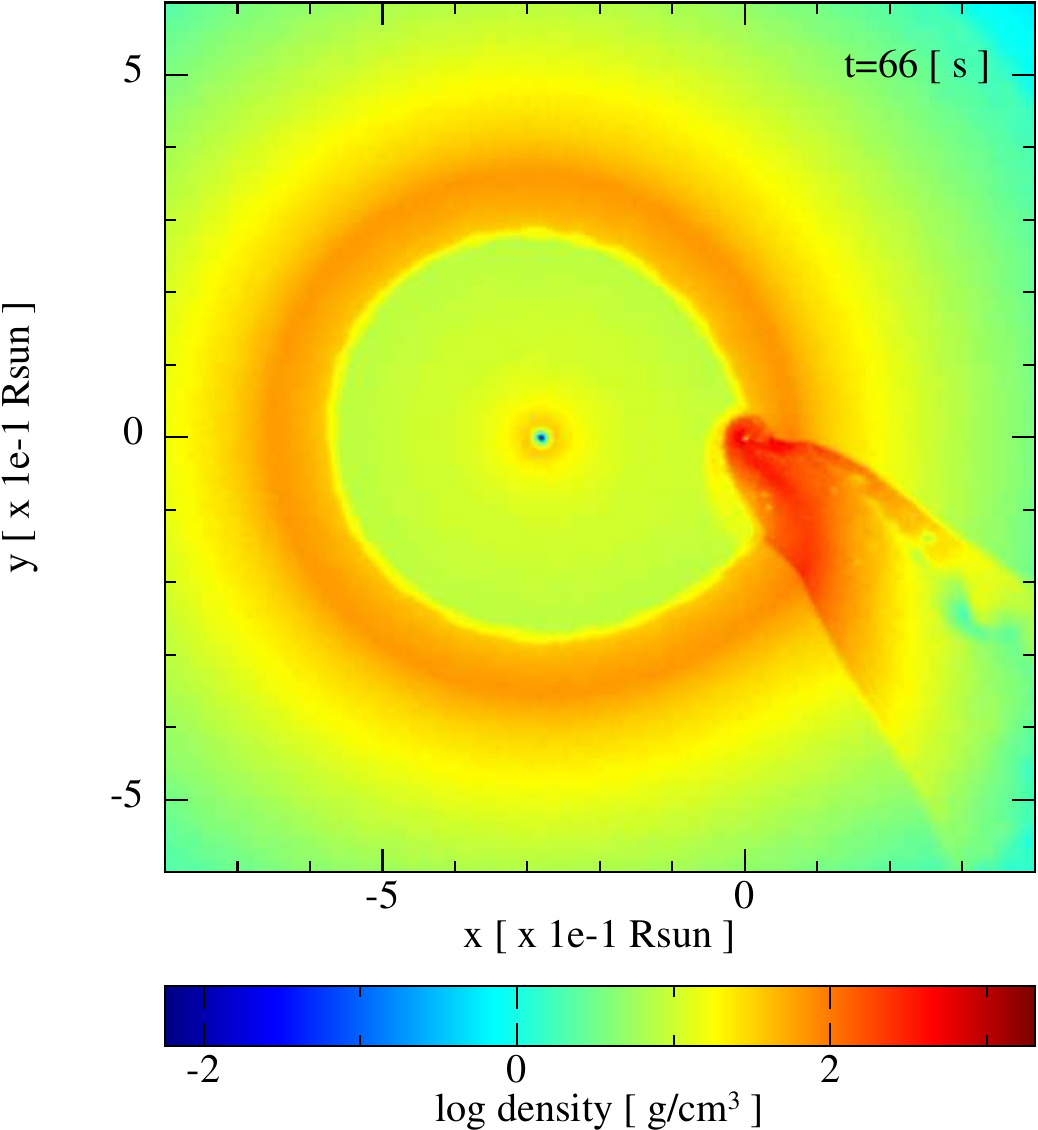} 
    
    \includegraphics[width=0.4\textwidth]{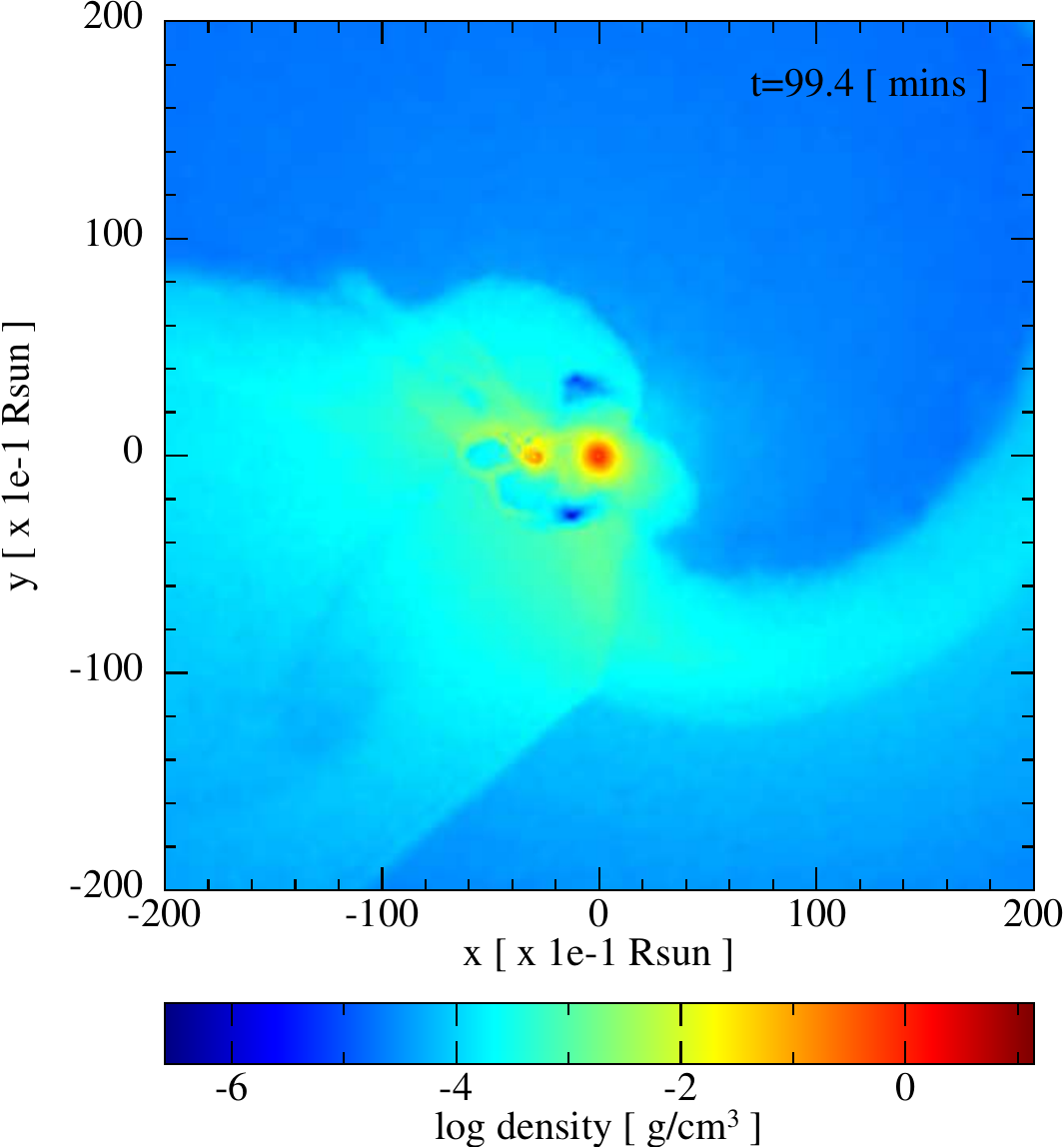} \includegraphics[width=0.4\textwidth]{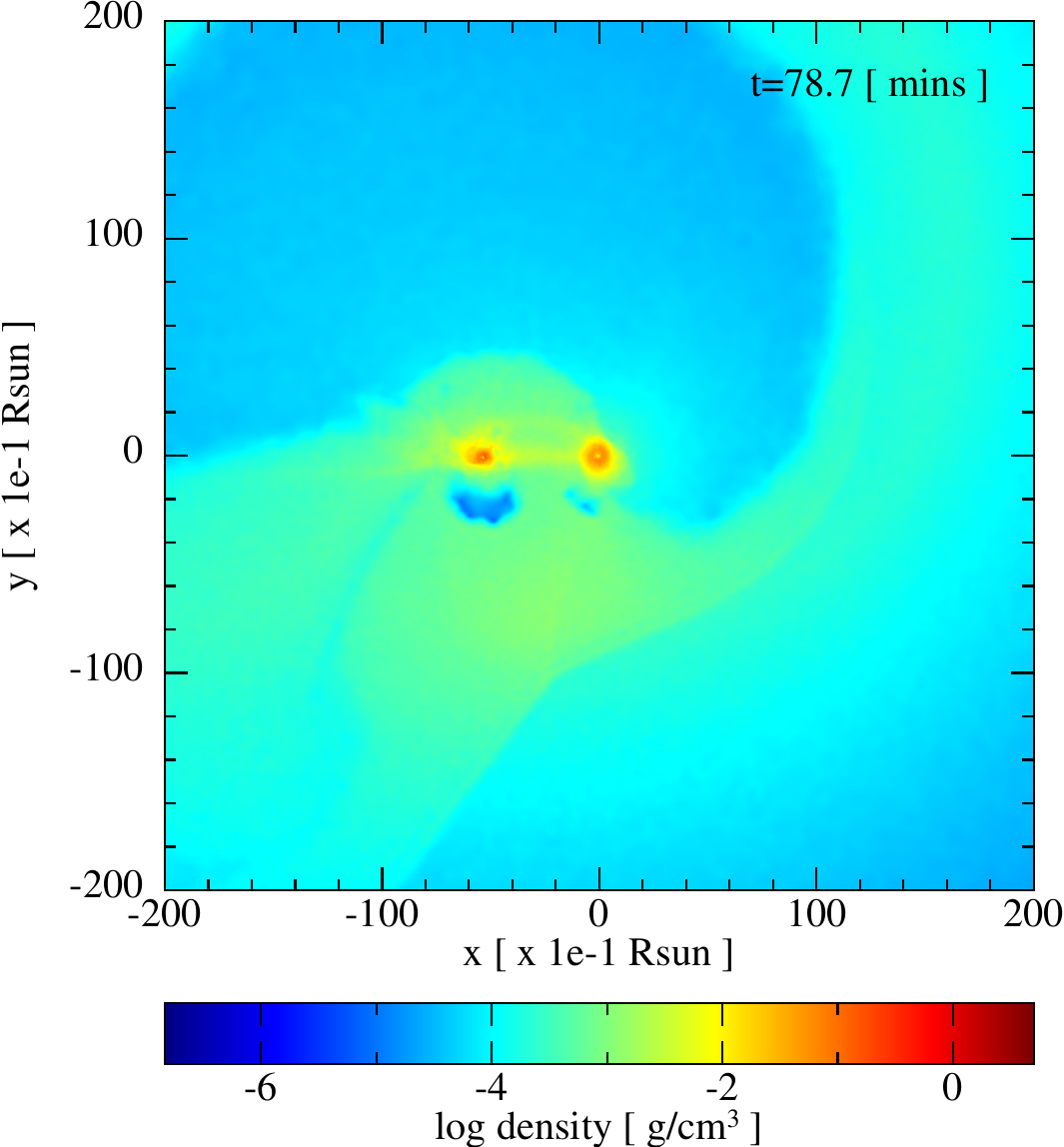} 
    \caption{  Snapshots of the SN ejecta mass density in the binary equatorial plane. The initial binary system is formed by a $2~M_\odot$ NS companion and a CO with $M_{\rm zams}=25~M_\odot$  in an orbital period of $P_{\rm orb,i}=5$~min (left panel) and $P_{\rm orb,i}=7.7$~min (right panel). The NS companion is at the origin of the coordinate system, and to its left, along the $x$-axis, it is the $\nu$NS. The material circularizing around the NS companion and a denser region around the $\nu$NS resulting from the fallback accretion can be seen in the upper panels. A disk also forms around the $\nu$NS in the lower panels. The system on the left remains bound after the SN explosion, while the system on the right becomes unbound.
    }
    \label{fig:snap_25Mzams}
\end{figure*}
%

\subsection{Initial binary period}\label{sec:Pinit}

The outcome of the BdHNe event in a binary system could be either an NS-NS, BH-NS, or BH-BH, depending on whether or not one or both stars collapse by the accretion of the SN ejected material. In the following, we aim to determine the initial parameter space (i.e., the initial binary period, the SN energy, the initial NS companion mass, and the progenitor of the CO) in which these outcome systems remain gravitationally bound.

Therefore, we start performing simulations increasing the initial binary period, keeping fixed the initial mass of the NS companion ($=2M_\odot$), the progenitor of the CO ($M_{\rm zams}=25M_\odot$), and the SN energy ($= 6.30\times 10^{50}$~erg). Figure~\ref{fig:Etot_Evol} shows the time evolution of the total energy, $E_{\rm tot}$ (see equation~\ref{eq:Energy_orb}) for some of these simulations. In this case, the pre-SN progenitor of the CO had a total mass of $M_{\rm CO}=6.8~M_\odot$, and after its gravitational collapse, it leaves a $\nu$NS with $M_{\rm \nu ns,i}=1.85~M_\odot$, ejecting about $M_{\rm ej} =4.99~M_\odot$ through the SN explosion ($M_{\rm CO}=M_{\rm \nu ns}+M_{\rm ej})$.  All the post-SN systems, $\nu$NS-NS companion, initially have  a positive total energy, $E_{\rm tot,i}$, given by:
 \begin{equation}\label{eq:Ebinary_i}
     \frac{ E_{\rm tot,i} }{|E_{\rm pre-SN}|} =\frac{M_{\rm \nu ns,i}M_{\rm ns,i} + M_{\rm CO}^2}{M_{\rm CO}(M_{\rm CO}+M_{\rm ns,i})}-2\frac{M_{\rm \nu ns,i}}{M_{\rm CO}}\, ,
 \end{equation}
 with $E_{\rm pre-SN}$ the binary system energy before the CO collapse:
\begin{equation}
     E_{\rm  pre-SN}=-\frac{1}{2}\frac{G M_{\rm CO}M_{\rm NS}}{a_{\rm orb, i}}\, .
\end{equation}
As the SN ejecta expands, the $\nu$NS and NS companion accrete mass and linear momentum, decreasing the total energy of the binary system. The system becomes gravitationally bound if the total binary energy evolves towards a negative value. Otherwise, it remains unbound.

Figure~\ref{fig:snap_25Mzams} shows snapshots of the SN ejecta mass density at the binary equatorial plane for two of the systems of Figure~\ref{fig:Etot_Evol}: one in which the outcome of the BdHne remains gravitationally bound (left panel with  $P_{\rm orb,i}=5$~min) and the other one in which it becomes unbound (right panel $P_{\rm orb,i}=7.7$~min). In both cases, at early times, the SN material that the NS companion gravitationally captures forms a kind of thick disk around the star before being accreted by it. At the same time, the innermost layers of the SN will produce a  fallback accretion onto the $\nu$NS. Later, the material circularizing around the NS companion is also attracted by the $\nu$NS, forming a disk around it. This second accretion episode produces a  peak in the $\nu$NS accretion rate, which is associated with the presence of its close companion \citep[see][for a more detailed description]{2019ApJ...871...14B, PhysRevD.106.083002}. 

\subsection{SN energy and NS companion initial mass}\label{sec:SNenergy}

\begin{table}
  \centering
\begin{tabular}{c|cc}
$E_{\rm sn}(10^{50}~{\rm erg} )$ & $\qquad b \qquad$ & $\qquad c\qquad$ \\ \hline
$6.30$ & $-3.153\pm 0.022$  & $5.219\pm 0.179$  \\
$5.67$ & $-3.38\pm 0.236$  & $4.803\pm 1.974$\\
$5.04$ & $-3.328\pm 0.137$&$0.195\pm 0.011$ \\
$4.73$ &  $-3.929\pm 0.118$ & $2.739\pm 0.972$\\
$4.41$ &$-5.748\pm 0.104$ &  $9.864\pm 0.846$\\
\hline
\end{tabular}
\caption{Parameters of the fitting relation given by Eq.~(\ref{eq:max_P}).}
  \label{tab:fit_x}
\end{table}
\begin{table}
  \centering
\begin{tabular}{c|c|c|c}
$M_{\rm ns}(M_\odot )$ & $P_0$ & $E_0$ & $\sigma$ \\ \hline
2.0 & $ 2.281\pm 0.434 $  &$ 8.073\pm 1.997$ & $3.582 \pm 2.130 $  \\
1.4 & $3.657 \pm 1.566 $ & $6.285 \pm 0.577 $ & $ 1.734 \pm 0.784 $\\
\hline
\end{tabular}
\caption{Parameters of the fitting relation given by Eq.~(\ref{eq:Pmax}).}\label{tab:fit_Esn}
\end{table}

\begin{figure}
    \centering
    \includegraphics[width=\columnwidth,clip]{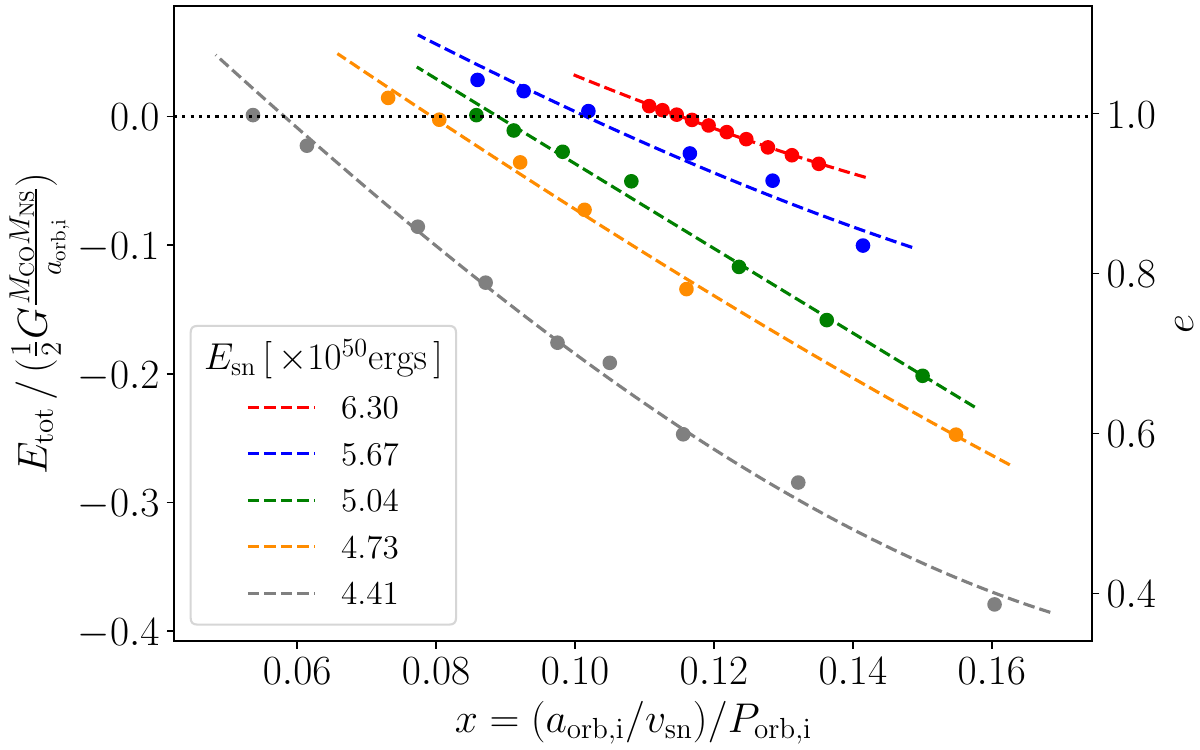}
    \caption{Final total energy (left axis) and eccentricity (right axis) of the binary system in function of $x = (a_{\rm orb,i}/v_{\rm sn})/P_{\rm orb,i} $ for different SN energy, $E_{\rm sn}$. The dashed lines correspond to the fitted relations (see Eq. \ref{eq:max_P}). The original binary comprises an evolved CO from a ZAMS progenitor of $M_{\rm zams}=25M_\odot$ and a $2M_\odot$ NS companion.}
    \label{fig:fit_25Mzams}
\end{figure}

We ran further simulations with different SN energies, $E_{\rm sn}$. For the initial setup, we scaled the kinetic and internal energy of the SPH particles by a constant factor $\eta$ (i.e., we multiplied the particle velocities by $\sqrt{\eta}$ and the internal energy by $\eta$). Figure~\ref{fig:fit_25Mzams} shows the total final energy of the system for these simulations as a function of $ x = (a_{\rm orb,i}/v_{\rm sn})/P_{\rm orb,i} $, where $v_{\rm sn}=\sqrt{ 2E_{\rm sn}/M_{\rm ej} }$, is the characteristic velocity of the SN ejecta. The variable $x$ gives the ratio between the characteristic time taken for the SN matter to arrive at the initial NS position and the initial orbital period. Thus, for $x$ close to zero, the mass loss from the SN event can be assumed instantaneous. With the results in Figure~\ref{fig:fit_25Mzams}, we have obtained an analytical  fit relation of the final total energy, i.e.,
\begin{equation}\label{eq:max_P}
    \frac{ E_{\rm tot,f} }{|E_{\rm pre-SN}| }\approx a + b x +c x^2\, ,
\end{equation}
The constants $b$  and $c$ are reported in Table~\ref{tab:fit_x} for some SN energies, while $a$ is given by Eq. (\ref{eq:Ebinary_i}), i.e.,
\begin{equation}
    a = \left( E_{\rm tot,f} / E_{\rm pre-SN} \right)_{x=0}.
\end{equation}
For instance, for a binary comprised of an evolved CO from a ZAMS progenitor $M_{\rm zams}=25M_\odot$ and a $2M_\odot$ NS companion, one obtains $a=0.294$.
 
As mentioned before, the remnant systems of a BdHNe event remain gravitationally bound when the total binary energy becomes negative, $E_{\rm tot,f}<0$. Then, the CO-NS initial system with the longest initial binary period, $P_{\rm orb, max}$, is the one that makes $E_{\rm tot,f} =0.0$ (black dotted horizontal line in Figure~\ref{fig:fit_25Mzams}). For the initial binary formed by a $2\, M_\odot$~NS  and a CO evolved from a progenitor with $M_{\rm zams}=25M_\odot$, this happens when $x=0.115\pm 0.001$, $0.102\pm 0.002$, $0.079\pm 0.002$, $0.072\pm 0.002$ and $0.053\pm 0.001$,  corresponding to $P_{\rm orb, max}=7.15$, $12.26$, $21.8$, $59.15$ and $102.84$~min, for $E_{\rm sn}=6.30$, $5.67$, $5.04$, $4.73$ and $4.41 \times 10^{50}$~erg, respectively. 

%
%

%
\begin{figure}
    \centering
    \includegraphics[width=0.99\columnwidth,clip]{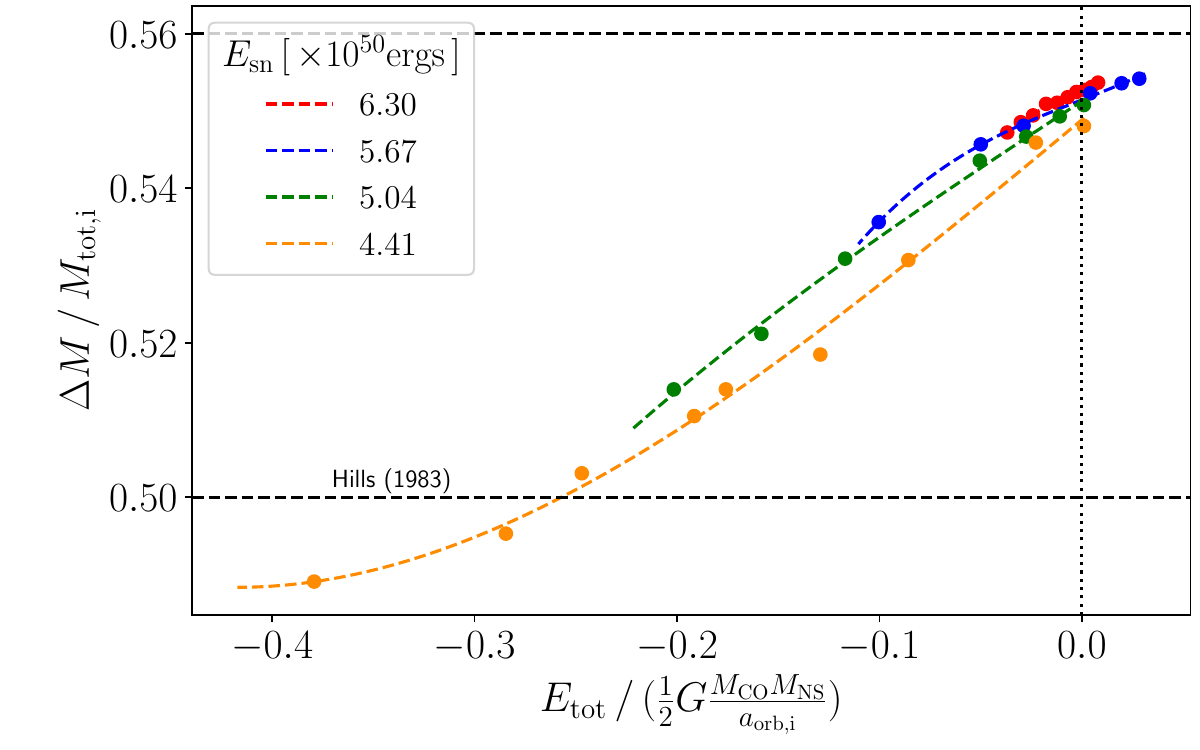}
    \caption{Mass loss (normalized to the initial binary mass) as a function of the final binary energy. The binary parameters are those of Fig.~\ref{fig:fit_25Mzams}.
    }
    \label{fig:Massloss_25Mzams}
\end{figure}

Usually, it is assumed that the binary system loses the material expelled by the SN explosion instantaneously. In this case, \cite{1983ApJ...267..322H} found that the binary orbital separation of the post-SN system is given by
\begin{equation}\label{eq:Hill_a}
    \frac{ a_{\rm orb, f} }{ a_{\rm orb,i}  }= \frac{M_{\rm tot,i} -\Delta M}{ M_{\rm tot,i} - 2a_{\rm orb,i} \Delta M /r }\, ,
\end{equation}
being $r$ the initial separation between the stars, $M_{\rm tot,i}$ the total mass of the pre-SN binary system and $\Delta M=M_{\rm tot,f}-M_{\rm tot,i}$, the change of mass between the post-SN and pre-SN binary system (for the BdHNe model $M_{\rm tot,i} = M_{\rm CO}+M_{\rm NS,i}$, and $M_{\rm tot,f}=M_{\rm ns,f}+M_{\rm \nu ns,f}$). 

From Eq. (\ref{eq:Hill_a}), it is concluded that a binary with a circular orbit will be disrupted after an SN event if more than half of its total initial mass is ejected in the explosion. Then, regardless of the initial orbital period or the SN energy, under the instantaneous mass-loss approximation, a BdHN should disrupt a system with a $6.8~M_\odot$ CO and a $2M_\odot$ NS since $\Delta M/M_{\rm tot,i}\approx 0.553-0.564$ (even if we account for the initial fallback accretion on the $\nu$NS, which is about $0.06-0.1 \,M_\odot$). Figure~\ref{fig:Massloss_25Mzams} shows the ratio between the mass change and the total mass of the initial binary system for the same simulations of Figure~\ref{fig:fit_25Mzams}. Most of the systems that remain gravitationally bound ($E_{\rm tot}<0)$ lose more than half of their initial total mass, in contrast to the traditional results based on the treatment by \citet{1983ApJ...267..322H}.

For the BdHNe model, Eq.~(\ref{eq:Hill_a}) does not hold because the assumption of instantaneous mass-loss breaks down \citep{2015PhRvL.115w1102F}. For these systems, the average time taken for the SN to reach the NS companion is more than $10\%$ of the initial binary period (see Fig.~\ref{fig:fit_25Mzams}). It should be noted that the slower innermost layers of the SN  take more time, while the faster outermost layers take less time. In addition, some of the mass ejected in the SN explosion is accreted by the $\nu$NS and the NS companion. This process reduces the mass loss and transfers linear momentum from the SN ejecta material to the stars.

\begin{figure}
    \centering
    \includegraphics[width=0.99\columnwidth,clip]{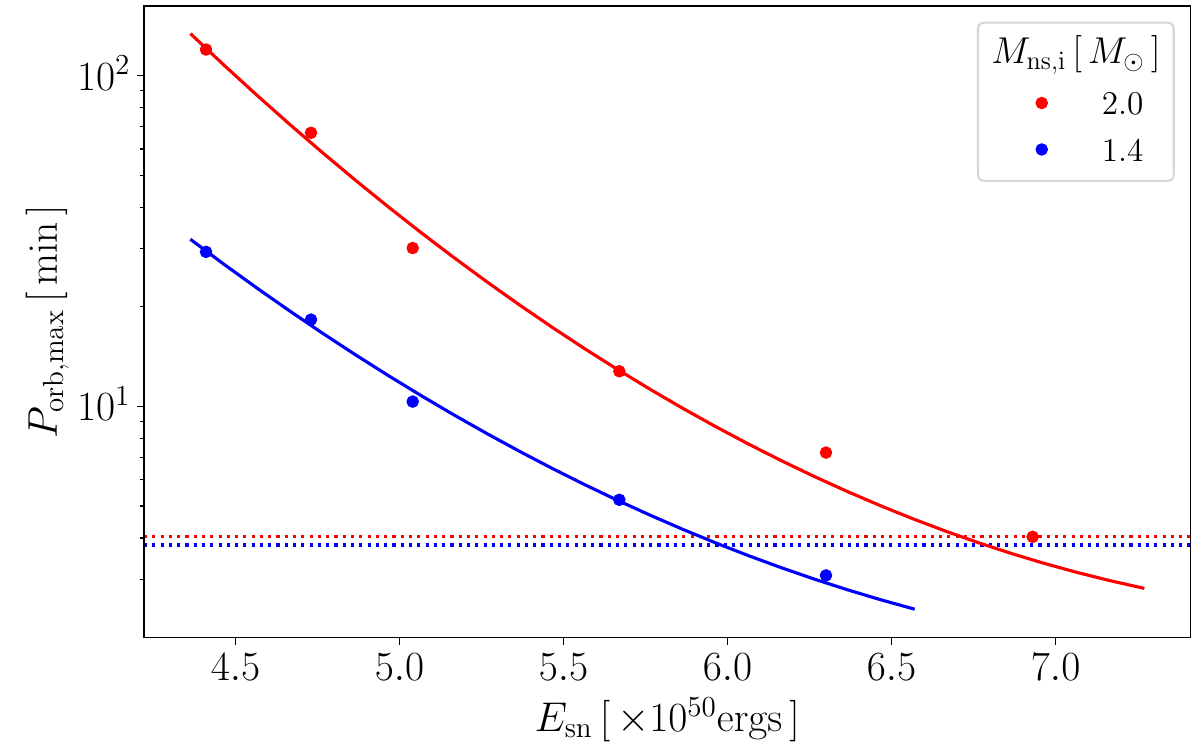}
    \caption{Maximum initial orbital period for which the binary after the BdHN event remains bound. The plot is given as a function of the SN energy. The CO star is that of the previous figures, and the initial NS companion mass is $1.4 M_\odot$ (blue) and $2 M_\odot$ (red). The horizontal dotted lines are the minimum orbital period of the system to avoid Roche-Lobe overflow before the SN explosion.
    }
    \label{fig:PorbMax}
\end{figure}

We run similar simulations (varying the initial binary period and the SN energy) for an initial  $1.4M_\odot$ NS companion. Figure~\ref{fig:PorbMax} shows  the relation between $P_{\rm orb,max}$ and $E_{\rm sn}$. The following function also fits this relation:
\begin{equation}\label{eq:Pmax}
     P_{\rm orb,max} = P_0 \exp\left[\frac{( (E_{\rm sn} - E_0)/(10^{50} { \rm erg}) )^2}{\sigma}\right],
 \end{equation}
where the parameters $P_0$, $E_0$ and $\sigma$ are in Table~\ref{tab:fit_Esn}. It is worth noting that there is a maximum SN energy that will lead to bound systems. Above that energy, all binaries are unbound regardless of the initial binary period. This energy is defined when the maximum orbital period to remain bound equals the minimum period to have no Roche-lobe overflow before the CO core collapses. The maximum SN energy is about $5.97\times 10^{50}$~erg and $6.64\times10^{50}$~erg for the $1.4~M_\odot$ and $2~M_\odot$ NS companion, respectively, with a CO with $6.8~M_\odot$.

\subsection{CO progenitor and SN geometry}\label{sec:COprog}

\begin{figure}
    \centering
    \includegraphics[width=\columnwidth,clip]{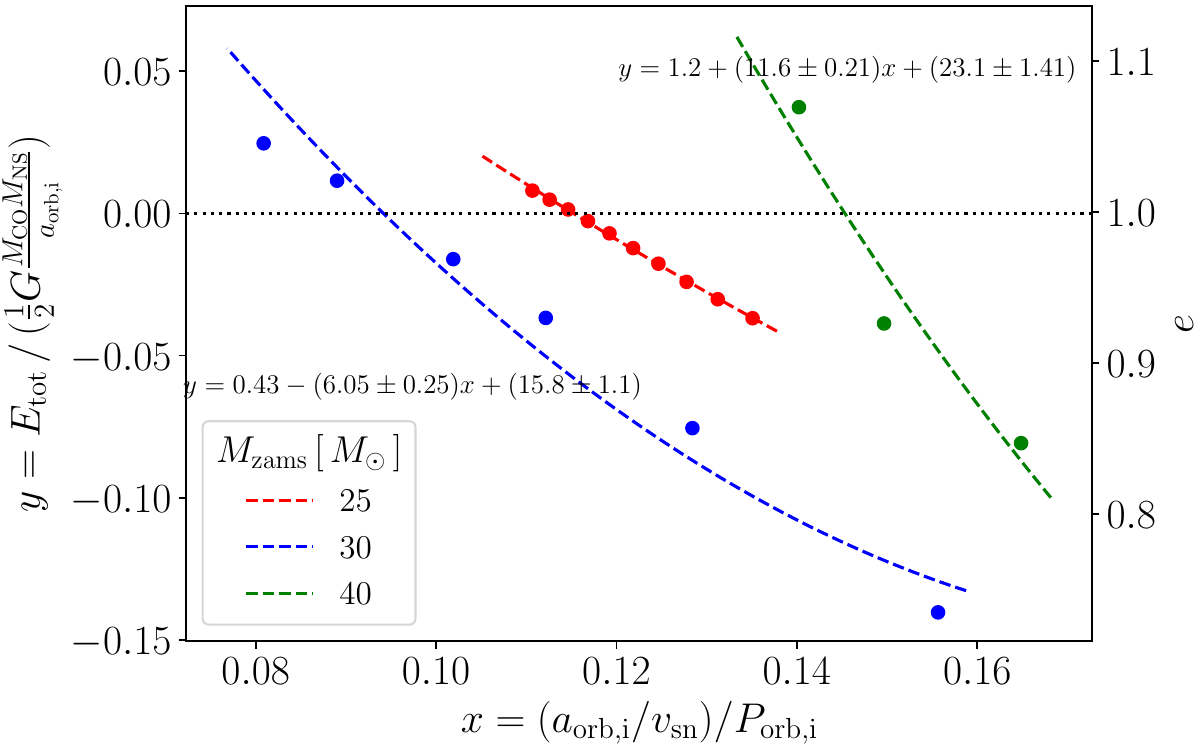}
    \caption{Same as Fig.~\ref{fig:fit_25Mzams} but for different ZAMS progenitors of the CO star.
    }
    \label{fig:Progenitor}
\end{figure}

We run simulations with more progenitors for the CO star. Figure~\ref{fig:Progenitor} shows the final energy of the binary as a function of the parameter $x$, with different progenitors for the CO and a $2M_\odot$ NS companion. The CO progenitor evolving from a progenitor with $M_{\rm zams}=30M_\odot$  has a total mass of $8.9M_\odot$ before the SN explosion. After its gravitational collapse, it leaves a $\nu$NS of about $1.75M_\odot$ and ejects about $7.13M_\odot$. The SN explosion total energy is about $6.54\times 10^{50}$~erg. While the progenitor with $M_{\rm zams}=40M_\odot$ has a total mass of $12.9M_\odot$ at the collapse moment, leaves a $\nu$NS of $1.85M_\odot$ and ejects $11.1M_\odot$ in the SN with a total energy of $1.56\times 10^{51}$~erg. All the systems should be disrupted after the SN event under the mass instantaneous assumption, since  $\Delta M/M_{\rm tot, i}\approx 0.65$ and $0.74$ for the $M_{\rm zams}=30$ and $40M_\odot$  CO progenitors, respectively. However, the SN explosion disrupts the binary for orbital periods longer than $26.5$ and $5.04$~min for the $30$ and $40M_\odot$ progenitors. 

\begin{figure}
    \centering
    \includegraphics[width=0.99\columnwidth,clip]{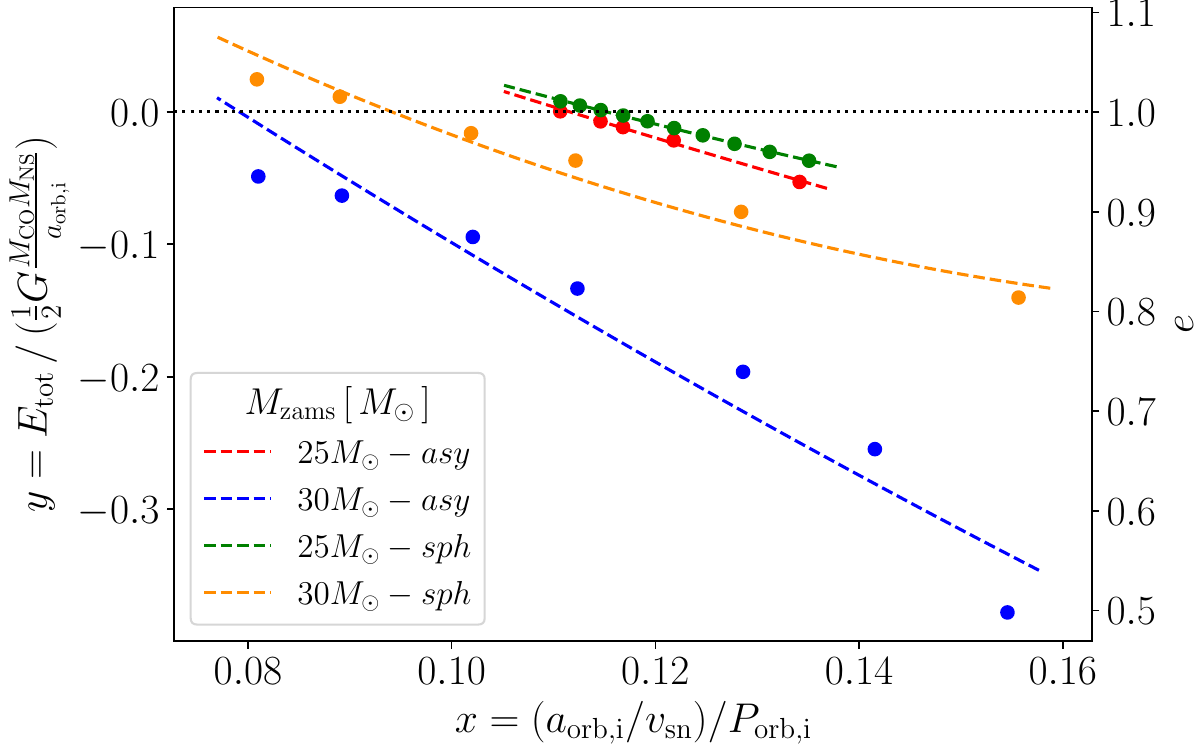}
    \caption{Same as Fig.~\ref{fig:fit_25Mzams} but for non-spherical symmetric SN explosion and different ZAMS progenitors of the CO star.}
    \label{fig:Esn_asys}
\end{figure}

Finally, we have also run simulations with a non-spherical SN explosion. For this, we modified the particle velocity, following \cite{2003ApJ...594..390H,2006ApJ...640..891Y}, to give a conical geometry to the SN ejecta \citep[see also][]{2019ApJ...871...14B}:
\begin{equation}
    V_{\rm in-cone}=f\left[\frac{1-f^2}{2}\cos\Theta+\frac{1+f^2}{2}\right]^{-1/2}V_{\rm sym}
\end{equation}
\begin{equation}
    V_{\rm out-cone}=\left[\frac{1-f^2}{2}\cos\Theta+\frac{1+f^2}{2}\right]^{-1/2}V_{\rm sym}\, .
\end{equation}
Here, $\Theta$ is the opening angle of the cone, and $f$ is the ratio of the velocities between the particles inside and outside the cone. The particles inside the code have the velocity $V_{\rm in-cone}$ and the particles outside it a velocity $V_{\rm out-core}$, while  $V_{\rm sym}$ is the radial velocity of the original explosion. Figure~\ref{fig:Esn_asys} shows the final energy of the binary system as a function of the parameter $x$ for spherical symmetry SN explosion and asymmetric ones. For these simulations, we used $f=2$ and $\Theta=20^{°}$ (this angle is measured from the $+z$-axis, normal to the initial binary system orbital plane). In general, the influence on the final binary energy is bigger for the $M_{\rm zams}=30M_\odot$ progenitor of the CO than for the $M_{\rm zams}=25M_\odot$ one.


\begin{figure}
    \centering
    \includegraphics[width=0.99\columnwidth,clip]{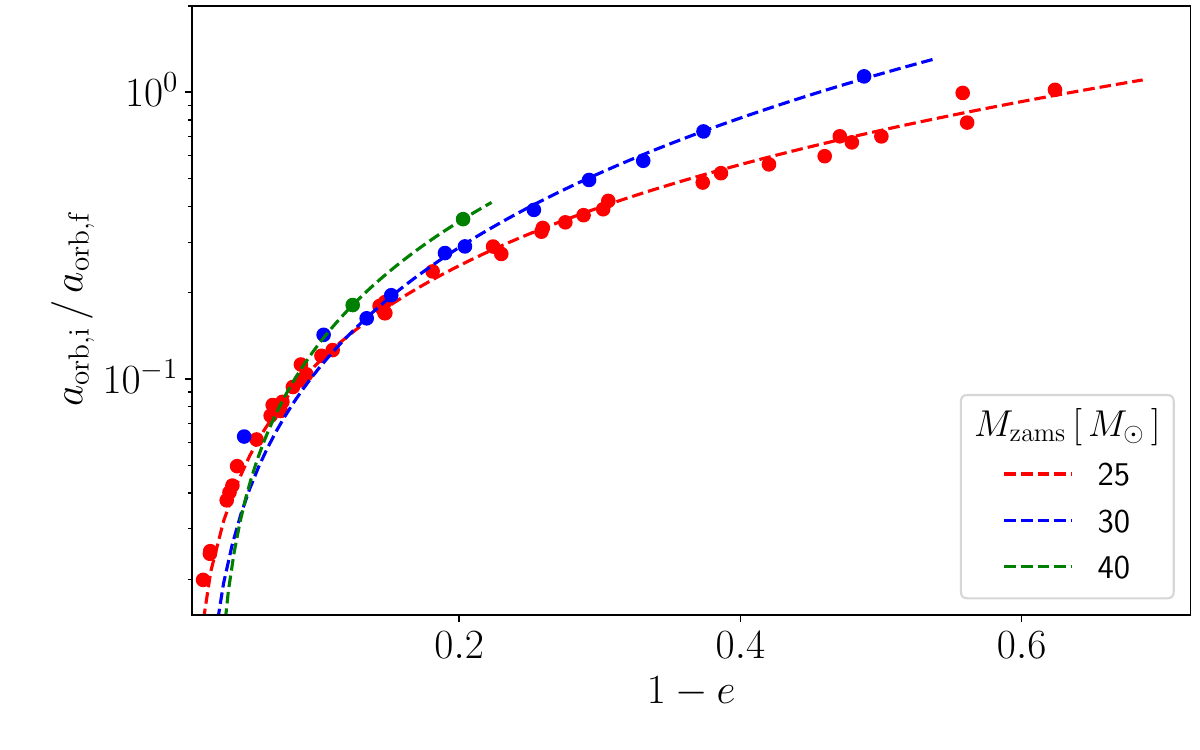}
    \caption{Ratio between the initial and final binary separation as a function of the final eccentricity for binary systems comprising a CO star (of three possible masses inferred from three ZAMS progenitors) and a $2M_\odot$ NS companion.}
    \label{fig:aorb_final}
\end{figure}

Finally, Fig.~\ref{fig:aorb_final} shows the relation between the eccentricity and the ratio between the bound system's initial and final binary separation. The relation between these quantities seems to depend on the CO mass progenitor and is independent of the SN kinetic energy. Also, the final binary system generally has a bigger eccentricity for larger binary separation. 

%

\section{Conclusions}\label{sec:clonclusions}

We have performed SPH simulations of the BdHNe under a wide range of the system's initial parameter space, varying both the initial binary separation, the mass of the supernova progenitor, and the characteristics of the explosion (both energy and asymmetry). We have established the conditions under which the binary system remains bound. We have given fitting formulas of the relevant parameters useful for further astrophysical analyses and applications. 

It has been particularly important for the bound systems computing the final binary orbit features such as the component masses, orbital period, eccentricity, and center-of-mass velocity. This information is essential to constrain the long-short GRB connection of the BdHN scenario, i.e., that the bound compact-object binaries left by BdHN events (NS-NS or NS-BH) become progenitors of short GRBs \citep{2015PhRvL.115w1102F, 2016ApJ...832..136R, 2018ApJ...859...30R}. For instance, the present simulations imply merger times by gravitational-wave emission in the range $10^4$--$10^9$ yr, which, using the systemic velocities $10$-$100$ km s$^{-1}$, can be used to obtain the traveled distances by the NS-NS and NS-BH binaries before the merging process and compare them with the difference of the measured long and short GRBs position offset in their host galaxies (Becerra et al., submitted; \citealp{2022ApJ...940...56F, 2023arXiv231012202N}).

The merger timescale also has implications for producing r-process elements from NS-NS mergers.  Galactic chemical evolution models have argued that such events alone can not explain all the r-process element production in the Milky Way. One argument focuses on the fact that merger times will delay the formation of r-process elements, so early times formation requires a second source~\citep[e.g.][]{2017ARNPS..67..253T,2019ApJ...875..106C}.  However, most of these studies assume a power-law distribution of merger times. For tight-orbit binaries, we find that the physics in our simulations predict a large fraction of bound, short-period binaries. Current recipes in population synthesis calculations do not include these physics effects. With a sizable population of short-period binaries with short ($<10^5$\,yr) merger times, NS-NS mergers may explain the observed r-process yields in the early universe.

Another argument for additional r-process sources is the observations of r-process elements in globular clusters and dwarf galaxies~\citep{2023arXiv231012202N}. Current distributions of merger times and post-formation systemic velocities would cause these mergers to escape these dwarf galaxies, preventing significant r-process enrichment of these galaxies. Here again, our results suggest that there could be a sizable population of very short merger time binaries that do not have time to move outside of the dwarf galaxy even if their systemic velocities are sufficient to escape the galaxy. Combining the work in this project with detailed population studies is needed to determine whether our improved physics can explain these observations.

\begin{acknowledgments}
 L.M.B. is supported by the Vicerrector\'ia de Investigación y Extensi\'on - Universidad Industrial de Santander Postdoctoral Fellowship Program No. 2023000359. The simulations were done on LANL HPC resources provided under the Institutional Computing program.  The work by CLF was supported by the US Department of Energy through the Los Alamos National Laboratory. Los Alamos National Laboratory is operated by Triad National Security, LLC, for the National Nuclear Security Administration of U.S.\ Department of Energy (Contract No.\ 89233218CNA000001)
\end{acknowledgments}


\end{document}